\documentstyle[prl,aps,epsf,multicol]{revtex}
\begin{document}
\draft
\title{Two-dimensional pattern formation in surfactant-mediated 
epitaxial growth}
\author{Bang-Gui Liu, Jing Wu, and E. G. Wang}
\address{Institute of Physics, Chinese Academy of Sciences, P. O. Box 603,
Beijing 100080, P. R. China}
\author{Zhenyu Zhang}
\address{Solid State Division, Oak Ridge National Laboratory, Oak Ridge, 
Tennessee 37831-6032, Department of Physics, University of Tennessee, Knoxville, Tennessee 37996}
%\date{\today}
\maketitle
\widetext
\begin{abstract}
The effects of a surfactant on two-dimensional pattern formation in
epitaxial growth are explored theoretically using a simple model,
in which an
adatom becomes immobile only after overcoming a large energy barrier as it
exchanges positions with a surfactant atom, and subsequent growth from such a
seed is further shielded.  Within this model, a fractal-to-compact 
island shape transition can be induced by either 
{\it decreasing} the growth temperature or {\it increasing} the 
deposition flux.  This and other intriguing findings are
in excellent qualitative agreement with recent experiments.
\end{abstract}

\pacs{PACS numbers: 68.55.Bd, 68.35.Bs, 68.55.Ln}

\raggedcolumns
\begin{multicols}{2}
\narrowtext

Because of stress effects, heteroepitaxial growth typically 
proceeds via the formation of three-dimensional (3D) islands, leading
to rough films.  However, it was discovered nearly a decade ago
that the use of a surfactant can lead to layer-by-layer
growth and drastically reduced film roughness \cite{Tromp}.  Since then,
much effort has been devoted to
the study of surfactant-mediated growth in both hetero-
and homo-epitaxial systems \cite{reviews}.  In these studies, it 
has been observed that the surfactant atoms can not only
modify the 3D growth mode, but often induce the formation
of fractal-like 2D islands \cite{behm,demiguel,voig,Hwang}.  
To date, little effort has been 
devoted to the understanding of the precise formation mechanisms 
for such fractal islands in the presence of surfactant atoms.
Such understanding is vitally important because the morphology and the
distribution of the 2D islands formed at submonolayer coverages
can severely influence the growth mode in the multilayer regime.

Two-dimensional pattern formation is itself an important area
of statistical physics. In their classic work, Witten and Sander
demonstrated that a fractal island can be formed when random 
walkers join a seed by hit-and-stick (without any relaxation)\cite{witten}.
More recently, 2D pattern formation within the context of 
dynamical island growth in submonolayer epitaxy has become 
the subject of intensive study, to a large extent advanced by the
capability of the scanning tunneling microscopy (STM) in characterizing
such islands\cite{BobHwang,Michely,Brune}.  These studies have firmly
established that islands can 
become more fractal-like if
the growth temperature is decreased at a given deposition flux, or 
the deposition flux is increased at a given growth 
temperature \cite{BobHwang,Michely,Brune,zhang1}.
However, most of the earlier studies of 2D pattern formation
had been focused on model homo- or hetero-epitaxial
systems without surfactants. Only very recently, have the effects
of Pb as a surfactant on the formation of 2D Ge islands grown
on Si(111) been studied systematically by 
Hwang et al.\cite{Hwang,Hwangis}.  They
found, most surprisingly, that the fractal-to-compact transition
is induced by lowering the temperature or by increasing the
deposition flux.  These observations are in clear contradiction
with traditional expectations, and the underlying physical 
mechanisms for such transitions are still unclear \cite{Hwang,Hwangis}. 
Michely et al. have also observed a compact to fractal transition
of Pt islands on Pt(111) by decreasing the deposition flux \cite{miche2},
possibly caused by the presence of the CO impurities \cite{miche3}. 
 In another experiment of
Sb-induced growth of C60 films on NaCl(100),
a compact-fractal-compact transition was observed by
increasing the temperature \cite{C60}.

In this Letter, we use a novel model to explore the 
effects of a monolayer of surfactant atoms
on 2D pattern formation in epitaxial growth.
The model contains a minimum number of key assumptions,
each based on sound physical grounds.  First, an adatom 
needs to overcome a rate-limiting potential energy barrier
in order to exchange positions with a surfactant atom
and become immobile.  Second, for other adatoms to join
such a seed and form a stable island, they still need
to overcome a repulsive potential energy barrier surrounding the seed. 
Third, only islands formed inside the surfactant layer are stable.
Our study of this simple model leads to various intriguing 
findings on both the morphology and the distribution
of the 2D islands formed under surfactant action.  
Most notably, a fractal-to-compact island transition
can be induced by either {\it decreasing} the growth
temperature or {\it increasing} the deposition flux.
We also obtain the characteristic dependences of the island
density as a function of temperature ($T$), flux ($F$), 
and coverage ($\theta$),
and rationalize our findings based on a simple physical picture
emphasizing the shielding effect on the incoming adatoms
by the surfactant atoms surrounding the islands.
Our findings are in excellent qualitative agreement with
the observations of Ge growth on 
Pb-covered Si(111),\cite{Hwang,Hwangis} and may 
find different degrees of applicability in other 
surfactant-induced growth systems 
as well \cite{behm,demiguel,voig,miche2,C60}. 

We start with an ideally flat substrate of material A, covered with 
a complete surfactant layer of material S. 
Atoms of a different 
type, B, are deposited onto the 
surfactant layer at a given deposition rate. 
We consider the case
where the coverage of S is sufficiently high, such that
the adatom islands, once formed, are always surrounded
by the surfactant atoms.
As our first study, we consider a simple model 
that catches the essential 
physics involved in the shape transitions but with a
minimum number of input parameters (hereafter referred as the first
model).  
Three elementary rate processes are emphasized in this 
model: diffusion ($dif$)
of a B-type atom on top of the surfactant
layer; a B-type adatom diving ($d$) from above to below the surfactant layer
(via place exchange with an S-type atom); and the aided diving ($ad$)
of a subsequent 
B atom to join the first one. We denote the activation barriers of these 
three processes by $V_{dif}$, $V_d$, and $V_{ad}$, respectively, and
the corresponding rates by $R_{dif}$, $R_d$, and $R_{ad}$, 
with $R=\nu\exp(-V/kT)$.  The three barriers satisfy the 
inequality chain $V_{dif} \ll V_{ad} < V_{d}$.  $V_{dif}$ is the smallest
because adatom diffusion is often significantly enhanced due to the
passivation of the surface by the surfactant layer \cite {kandel1}.  
$V_{d}$ is the
largest, making it the rate-limiting process for eventual formation 
of a stable island.  The first inequality reflects the fact that 
there exists an effective repulsive wall surrounding the seed 
atom or an island formed underneath the surfactant layer.
The existence of such a repulsive potential to the incoming
adatoms due to the presence of the surfactant atoms surrounding an
island has been proposed previously \cite{Markov,kandel1},
and its effect on the island density has been explored very
recently \cite{kandel2,Hwang}.
An isolated adatom can hop with the rate $R_{dif}$,
or exchange down with the rate $R_d$.
Here for simplicity, the in-plane mobility of the B atoms 
underneath the surfactant layer is taken to be negligible.  
We also ignore the reverse exchange process in which a B-type atom 
resurfaces to the top of the surfactant layer, corresponding 
to the case where a B atom strongly favors the underneath site.  
On the other hand, islands formed on top of the surfactant
layer can still dissociate, and are therefore unstable. 
When a B atom hops to a site which has $n_d$ static
B-type nearest neighbors, it remains stuck there 
until it exchanges down with the rate $n_dR_{ad}$.
Because the activation barrier for this process must be in between $V_{ad}/n_d$
and $V_{ad}$, for simplicity we choose the rate $n_dR_{ad}$ to take into account
the effect of the $n_d$ neighboring static atoms without introducing
another parameter at this stage.
Using the definition of classical nucleation theory \cite{venables}, 
we actually have two critical
island sizes: $i^*=\infty$ and $i^*=0$ for the upper layer and 
the lower layer, respectively. 
In contrast to the earlier irreversible "hit-and-stick" \cite{witten}
or the "hit-stick-relax" model \cite{zhang1}, the stable islands
in the present study consist only of down-exchanged atoms. 
Overall, the current model is consistent with the
fact that the binding energy between A and B is typically much larger than
that between A and S.\cite{Schroeder}  Later we will show that considerations
going beyond this first model do not change the main qualitative
features of the present study.

We primarily use kinetic Monte Carlo (KMC) simulations to study this model;
later we also briefly describe the main results from 
rate equation analysis.  The KMC simulations were carried
out on a square $200 \times 200$ lattice, though
simulations using a lattice of triangular geometry yield qualitatively
similar conclusions.
We take a small diffusion barrier $V_{dif}=0.59$ eV, reflecting the
fast adatom diffusion on top of the surfactant layer. 
The barriers of the exchange (diving) and the aided exchange processes are 
taken as $V_d=0.90$ eV and $V_{ad}=0.82$ eV, respectively. 
The attempt frequency is uniquely chosen to be
$\nu = 4.1671\times 10^{10} T$, with $T$ given in degree K.

The temperature dependence of the island shapes
obtained at $F=0.005$ ML/s 
and $\theta=0.1$ ML is shown in Fig. 1. 
At 300 K, the islands are typically compact (Fig. 1a);
at 340 K, the islands 
are typically fractal-like. The transition from compact to fractal patterns
takes place approximately at 315 K. Fig. 1b shows the pattern at 
310 K, just below the transition temperature; the islands are still compact,
though there are some ramified structures in the outer part of the islands.
Fig 1c is the pattern at 320 K, just above the transition; 
here the islands are predominantly fractal-like.

The intriguing temperature dependence described above
can be understood by considering the 
shielding effect of the adatoms stuck around the edge of a nucleation
seed or an island of down-exchanged atoms in the sublayer.  At high
temperatures, such surrounding adatoms can easily dive into the sublayer
at their initial points of sticking; once they manage to
exchange into the sublayer, their mobility is severely limited, making
the whole situation very similar to the classic hit-and-stick 
diffusion limited aggregation \cite{witten}.
On the other hand, at lower temperatures,
such stuck adatoms and the surfactant atoms surrounding them
effectively block incoming adatoms from reaching a
seed atom or an island in the sublayer.  Therefore, 
these incoming adatoms have
a chance to leave their initial points of sticking, and after some 
random walking can restick at different points of the same island.  
Such processes effectively lead to relaxation around the edge of 
an island, resulting in more compact island morphology \cite{zhang1}.

Figure 2 shows an interesting nonmonotonous dependence 
of the island density with the temperature.
The minimum in island density is located right at the
temperature at which the compact-to-fractal transition
in island morphology has been observed.
We note that the temperature dependence 
shown here is similar to that obtained 
previously by Meyer and Behm\cite{meyer}, but the two cases
differ in physical origins. In their case, islands formed either
by nucleation of two mobile adatoms or by meeting of
one mobile and one trapped adatom are both stable, and the minimum in the 
island density as a function of the temperature
is associated with the transition from
the nucleation-dominant to exchange-prominent region. 
In our case, islands formed on top of the surfactant layer
are unstable, and the rate-limiting process for the nucleation
of a stable island is the diving of an adatom into the sublayer.
Therefore, our system is always in the exchange-prominent region.
Nevertheless, when the temperature is low, a stable seed 
atom in the surfactant layer may not necessarily grow into a stable island
because of the effective shielding of the incoming adatoms. But those
seeds which manage to grow into islands will grow even faster as
their sizes increase.  The decrease in island density with temperature
is caused by the increased mobility of 
the adatoms in searching for such islands.
On the other hand, after the transition temperature, the shielding
effect is very weak, and every seed atom is likely to
grow into a stable island.  The island density increase with temperature
reflects the enhanced rate in creating such seeds.

Figure 3 displays the island patterns obtained 
at different deposition rates. Here the
growth temperature
is fixed at $T=300$ K and the coverage is again at 0.1 ML. Fig. 3a is an ideal
fractal pattern at the flux of 0.0001 ML/s. Fig. 3b is still a fractal pattern
at the flux of 0.001 ML/s, though there are islands that are more compact.
In Fig. 3c the flux is 0.0025 ML/s and the island shape 
already becomes compact. In Fig. 3d the islands are ideal compact
patterns, obtained at the flux of 0.028 ML/s.  Because the shielding effect
increases with flux, the flux-induced fractal to compact
transition is also driven by
the shielding effect, but here the transition is from strong shielding
to weak shielding when the flux increases. 

Figure 4 shows the flux dependence of the stable island density, $N_s$,
 obtained at $T=300$ K
and $\theta=0.1$ ML. 
The curve can be divided into three regimes: the low-flux fractal regime,
where the dependence is very weak;  
the intermediate flux regime, where a scaling law can be well defined 
($N_s\sim F^{\beta}$ with $\beta = 0.40$);
and the high-flux compact regime, where the island density has
saturated.  When the maximal saturation island density 
(obtained at different coverages) is plotted
as a function of flux, a much larger scaling exponent is obtained
($\beta = 0.7$).
%, but it is clearly smaller than the experimental value for
%Ge/Si(111)-Pb\cite{Hwangis}.

We have also carried out a limited rate equation analysis of the
above model.
In this approach, we introduce the island perimeter $L_d$ and 
parameterize the area of an island by $S_d=pL_d^q$, 
where $p$ is a constant and $q$
is the dimension of the islands \cite{zhang1}.  For compact islands
we have $q=2$ and for
fractal patterns $1<q<2$.
In the low-coverage limit, 
we have the following rate equations:
\begin{equation}\label{1}
\frac{d}{dt}n_a=F\theta (t_0-t)-\alpha_dn_a-\alpha_bn_aN_s(L_d-n_b)
\end{equation}
\begin{equation}
\frac{d}{dt}(N_sn_b)=-\alpha_en_bN_s+\alpha_bn_aN_s(L_d-n_b)
\end{equation}
\begin{equation}
\frac{d}{dt}N_s=\alpha_dn_a
\end{equation}
where $n_a$ is the density of the movable active atoms,
 $n_b$ is the number of edge atoms per stable island, 
$\alpha_d$ and $\alpha_e$ can be taken as $R_d$ and $R_{ad}$, and $\alpha_b$
is the capture constant. We have introduced the
 step function $\theta{}(t_0-t)$ 
to reflect the fact that the STM imaging was typically
some time after 
the deposition time $t_0$.\cite{Hwang,Hwangis}
 Letting $N_a$ be the total number of
active atoms, one has $N_a=n_a+N_sn_b$.
When the flux is low or the temperature is high, 
very few adatoms remain active
at the end of deposition.
The island patterns can then be determined mainly by 
the $t<t_0$ region. In such situations,
a steady-state
approximation can be made for $N_a$, 
similar to what has been done previously
\cite{kandel1,Markov,kandel2,bales}. Then a scaling law of
island density can be derived: $N_s\sim F^{n_F}\theta^{n_{\theta}}$,
where $n_F=0$, as suggested in the low-flux regime shown in
Fig. 4; and $n_{\theta}=(q-1)/(2q-1)$. The constant $q$ can be
easily determined from the island shapes 
only in the limit of very low flux or very high temperature. 
In the compact and intermediate regimes, both the $t<t_0$
and $t>t_0$ regions should be considered. It is then more
involved to obtain a simple analytical scaling law \cite{Wu}.

The temperature- or flux-induced fractal-compact transition in the
island shapes predicted within the present model provides,
on a qualitative level, the theoretical basis for the
transitions observed 
in Pb-induced growth of Ge on Si(111) \cite{Hwangis}.
Because there are only three parameters in our first model, 
the agreement in the main features between theory and experiment  
should be viewed excellent. 
In particular, the shielding effect emphasized in the present
model plays the essential role in causing the shape transition.

We have expanded the range of applicability of our simple model
to various more realistic growth systems in several aspects,
such as KMC simulations on a triangular lattice
and variations of the three basic model parameters.
We have also considered physical effects beyond the first 
model, including the binding energy of the
adatoms in an island formed above the surfactant
layer, detachment of the adatoms trapped around the edge 
of a stable island, and the possibility of simultaneous exchange of
 multi-adatoms \cite{Hwang,Hwangis}.
These lattice and parameter variations, as well as
the improvement beyond the first model,
do not change the central qualitative findings of the present
work, namely, the counter-intuitive fractal-compact transition
caused by the temperature or the flux. Of course, one
should expect that, on a quantitative level, the island
density will depend differently on the temperature, flux,
and the coverage \cite{Wu}.  We should also note that the fractal-compact
transitions predicted here are expected to be observable even if
the stable islands are formed on top of the surface but surrounded
by a sufficiently high coverage of impurity atoms, as long as
those impurity atoms can effectively hinder the growth of the
stable islands by shielding.  Furthermore,
the phenomena are not limited to heteroepitaxial growth only:
Even in surfactant-induced homoepitaxy, similar phenomena are
likely to occur if the shielding effect associated with the
impurity or surfactant atoms is sufficiently effective.  

In summary, we have theoretically explored the effects of a monolayer 
of surfactant atoms on two-dimensional pattern formation in epitaxial 
growth by using a simple but physically sensible model. 
We find that a fractal-to-compact island transition
can be induced by either {\it decreasing} the growth temperature or
{\it increasing} the deposition flux.  Furthermore, 
the flux and temperature dependence
of the island density on the fractal side is very different from that 
on the compact side.  The counter-intuitive
predictions on the island morphological evolution
can be rationalized based on the shielding effects.
Our findings on the shape transitions 
are in excellent qualitative agreement
with recent observations, while the predicted nonmonotonic temperature
dependence of the island density is yet to be confirmed
in future experimental studies.

We thank I.-S. Hwang and T. T. Tsong for stimulating discussions,
and R. J. Behm and J. J. de Miguel for helpful correspondences.
This research was supported by the Chinese Natural
Science Fundation (Grant No. 19810760328), by Chinese State Key Project
of Basic Research on Rare Earth, by
Oak Ridge National Laboratory, managed by
Lockheed Martin Energy Research Corp. for the
U.S. Department of Energy under Contract No. DE-AC05-96OR22464,
and by the U.S. National Science Foundation (Grant No. DMR-9705406).

\newpage
\begin{figure}
\epsfxsize=0.5\textwidth
\epsfbox{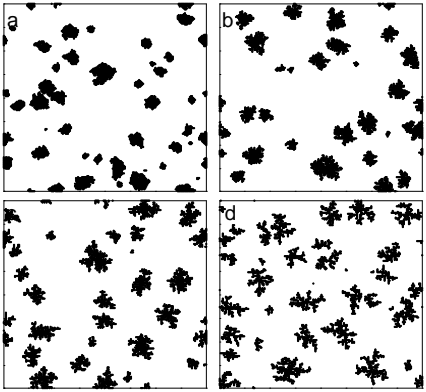}
\caption{Island shapes 
obtained on a $200\times{}200$ lattice with a constant
deposition flux of 0.005 ML/s and a constant coverage
of 0.1 ML, but at four different temperatures:  
(a) 300 K; (b) 310 K; (c) 320 K; and (d) 340 K. 
The compact-to-fractal transition takes place approximately at 315 K.}
\label{fig1}
\end{figure}

\begin{figure}
\epsfxsize=0.45\textwidth
\epsfbox{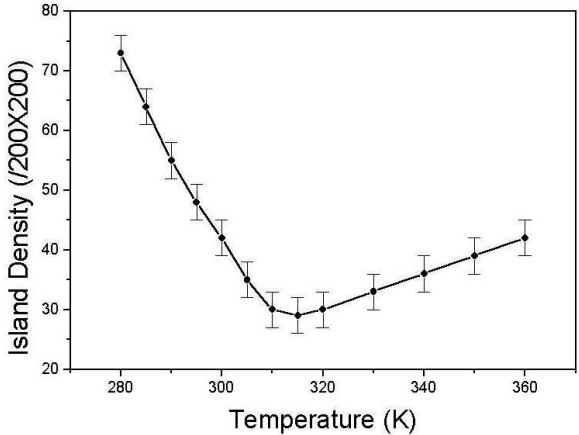}
\caption{Island density as a function of temperature,
with the flux fixed at 0.005 ML/s and the coverage fixed at 0.1ML. 
The temperature at which the density is a minimum 
is the same temperature for island shape transition.}
\label{fig2}
\end{figure}

\begin{figure}
\epsfxsize=0.5\textwidth
\epsfbox{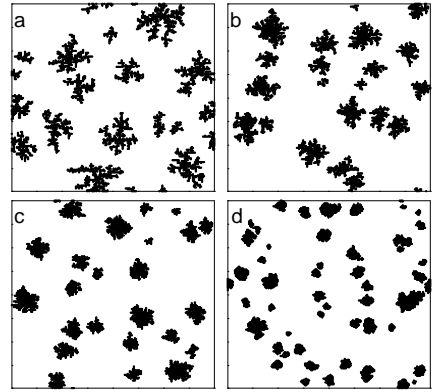}
\caption{Flux dependence of the island shapes obtained
 on a $200\times{}200$ lattice at a constant temperature
of 300 K and a constant coverage of 0.1 ML,
but with four different fluxes:
(a) 0.0001 ML/s; (b) 0.001 ML/s; (c) 0.0025 ML/s;
and (d) 0.028 ML/s. 
The shape transition takes place between
0.001 ML/s and 0.0025 ML/s.}
\label{fig3}
\end{figure}

\begin{figure}
\epsfxsize=0.45\textwidth
\epsfbox{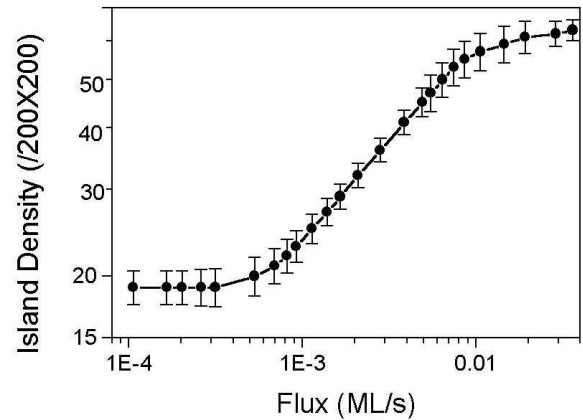}
\caption{Island density as a function of flux.
The temperature is fixed at 300 K and the coverage fixed at 0.1 ML. 
There exists an approximate scaling law in the intermediate flux
region.}
\label{fig4}
\end{figure}

\end{multicols}

\end{document}